\documentclass[conference]{IEEEtran}
 
\usepackage{cite}

\ifCLASSINFOpdf
   \usepackage[pdftex]{graphicx}
\else
 \usepackage[dvips]{graphicx}
\fi

\usepackage[cmex10]{amsmath}

\usepackage{amsfonts,amssymb}

\usepackage{breqn}
\allowdisplaybreaks

\usepackage{hyperref}

\usepackage{siunitx}

\makeatletter
\let\old@ps@headings\ps@headings
\let\old@ps@IEEEtitlepagestyle\ps@IEEEtitlepagestyle
\def\psccfooter#1{%
    \def\ps@headings{%
        \old@ps@headings%
        \def\@oddfoot{\strut\hfill#1\hfill\strut}%
        \def\@evenfoot{\strut\hfill#1\hfill\strut}%
    }%
    \def\ps@IEEEtitlepagestyle{%
        \old@ps@IEEEtitlepagestyle%
        \def\@oddfoot{\strut\hfill#1\hfill\strut}%
        \def\@evenfoot{\strut\hfill#1\hfill\strut}%
    }%
    \ps@headings%
}
\makeatother

\usepackage{url}

\usepackage[]{fancyhdr} % 
\newcommand{\changefont}{\fontsize{9}{9}\selectfont}
\usepackage{amssymb}
\usepackage{nomencl}
\makenomenclature

\usepackage{etoolbox}
\renewcommand\nomgroup[1]{%
  \item[\bfseries
  \ifstrequal{#1}{S}{Sets/Indices}{%
  \ifstrequal{#1}{P}{Parameters}{%
  \ifstrequal{#1}{V}{Variables}{}}}%
]}
\IEEEoverridecommandlockouts
\begin{document}

\title{Operating Envelopes under Probabilistic Electricity Demand and Solar Generation Forecasts}

\author{
\IEEEauthorblockN{Yu Yi, Gregor Verbi\v{c}}
\IEEEauthorblockA{The University of Sydney\\
NSW, Australia\\
\{yu.yi, gregor.verbic\}@sydney.edu.au}
}


%Telephone: (800) 555--1212, Fax: (888) 555--1212}

% <-this % stops a space


%\lhead{11TH BULK POWER SYSTEMS DYNAMICS AND CONTROL SYMPOSIUM, JULY 25-30, 2022, BANFF, CANADA}
%\rhead{1}

%\fontfamily{phv}\fontseries{b}\fontsize{9}{11}\selectfont

\maketitle
\thispagestyle{fancy}
\pagestyle{fancy}

%\thispagestyle{fancy}
%\pagestyle{fancy}

\begin{abstract}
The increasing penetration of distributed energy resources in low-voltage networks is turning end-users from `consumers' to `prosumers'. However, the incomplete smart meter rollout and paucity of smart meter data due to the regulatory separation between retail and network service provision makes active distribution network management difficult. Furthermore, distribution network operators oftentimes do not have access to real-time smart meter data, which creates an additional challenge. For the lack of better solutions, they use blanket rooftop solar export limits, leading to suboptimal outcomes. To address this, we designed a conditional generative adversarial network (CGAN)-based model to forecast household solar generation and electricity demand, which serves as an input to chance-constrained optimal power flow used to compute fair operating envelopes under uncertainty.

%As the increase of distributed energy resources (DER) installation on the residential areas in Australia, the users in low-voltage networks are changing their identities from 'consumer' to 'prosumer'. However, the incomplete and invisible smart meter data disrupt the schedule and operation on the power grid. Moreover, the redundant local injection push the electricity network to the edge of the limitations. To address these gaps, a conditional generative adversarial network (CGAN)-based model is designed and trained for the probabilistic forecasts of household solar generation and loads. Based on the probabilistic forecasts, we generated the nodal operating envelopes with the forecasting residuals using a chance-constrained AC optimal power flow (CC AC OPF) model.The performance of the probabilistic forecast methodology in the stochastic power flow model is evaluated in the case studies.
\end{abstract}


% ----- Nomenclature ------------------------------------------
% Sets/Indices
\nomenclature[S]{$\mathcal{N},i$}{Set/Index of nodes in network}
\nomenclature[S]{$\mathcal{L},ij$}{Set/Index of transmission lines}
\nomenclature[S]{$\mathcal{T},t$}{Set/Index of time slots}
\nomenclature[S]{$\mathcal{X}^{\text{p}},x^{\text{p}}$}{Set/Index of prosumer decision variables}
\nomenclature[S]{$\mathcal{X}^{\text{n}},x^{\text{n}}$}{Set/Index of network decision variables}
\nomenclature[S]{$\mathbb{P}$}{Probability of a stochastic variable remaining within limits}
\nomenclature[S]{$S_r$}{Set of the real training data}
\nomenclature[S]{$S_r$}{Set of the generated sample from generator network}

% Variables
\nomenclature[V]{$p^{\mathrm{exp}}_{i,t}$}{Nodal exporting power}

\nomenclature[V]{$\tilde{p}^{\mathrm{d}}_{i}$}{Stochastic household demand}
\nomenclature[V]{$\tilde{p}^{\mathrm{pv}}_{i}$}{Stochastic PV generation}
\nomenclature[V]{$\widehat{p}_{i}^{\mathrm{d}}/\widehat{p}_{i}^{\mathrm{pv}}$}{Household demand/PV generation point forecast}
\nomenclature[V]{$\varepsilon^{\mathrm{d}}_{i}/\varepsilon^{\mathrm{pv}}_{i}$}{Household demand/PV generation forecasting residuals}

\nomenclature[V]{$\tilde{p}^{\mathrm{b}}_{i}$}{Charging power of battery}
\nomenclature[V]{$\tilde{e}^{\mathrm{b}}$}{SOC of battery}
\nomenclature[V]{$b_{t}$}{Binary variable to decide charging/discharging status of battery}

\nomenclature[V]{$\tilde{v}_{i}$}{Square of voltage of node $i$}
\nomenclature[V]{$\tilde{\ell}_{i}$}{Square of current of node $i$}
\nomenclature[V]{$\tilde{p}_{ij}/\tilde{q}_{ij}$}{Active/reactive power injecting from bus $i$ to $j$}
\nomenclature[V]{$\tilde{s}_{ij}$}{Apparent power injecting from bus $i$ to $j$}

\nomenclature[V]{$\widehat{y_{t}}$}{Point forecasts}
\nomenclature[V]{$\varepsilon_{t}$}{Forecasting residuals}

\nomenclature[V]{$z$}{Random noise input of the CGAN model}
\nomenclature[V]{$L_{G}/L_{D}$}{Loss of the generator/discriminator}

% Parameter
\nomenclature[P]{$r_{ij}/x_{ij}$}{Resistance/reactance of transmission line $ij$}
\nomenclature[P]{$z_{ij}$}{Impedance of transmission line $ij$}
\nomenclature[P]{$\overline{p}^{\mathrm{exp}}$}{Maximum nodal exporting power}
\nomenclature[P]{$\overline{p}^{\mathrm{pv}}$}{Maximum PV generation}
\nomenclature[P]{$\underline{p}^{\mathrm{b}} /\overline{p}^{\mathrm{b}}_{t}$}{Minimum/Maximum charging power of battery}
\nomenclature[P]{$\underline{e}^{\mathrm{b}}/\overline{e}^{\mathrm{b}}$}{Minimum/maximum soc of battery}
\nomenclature[P]{$\eta$}{Charging efficiency of battery}
\nomenclature[P]{$\text{cos}\varphi$}{Power factor}
\nomenclature[P]{$\xi^{\text{v}}/\xi^{\ell}/\xi^{\text{p}}$}{Violation probabilities of chance constraints} 

\nomenclature[P]{$\mathbf{X}_{t}$}{Input factors of the CGAN modal}
\nomenclature[P]{$\mathbf{C}_{t}$}{Condition input factors of the CGAN model}
\nomenclature[P]{$\theta_{G}/\theta_{D}$}{Parameters of generator and discriminator network}

\nomenclature[V]{$\underline{\bullet}/\overline{\bullet}$}{Minimum and maximum limitations for variables}
\nomenclature[V]{$\tilde{\bullet}$}{Stochastic Variables}
\nomenclature[V]{$\widehat{\bullet}$}{Point forecast}

\printnomenclature

% \fi
\IEEEpeerreviewmaketitle

\section{Introduction}
%With the diverse integration of behind-the-meter distributed energy resources (DERs), including rooftop photovoltaic systems (PV), battery storage, and flexible loads, the way of electricity generation and consumption is changing. The increasing installed capacity of PVs and the improving small-scale energy storage technologies at low-voltage networks will push the electricity network on the verge of the technic limits. Furthermore, the solar generation and demand forecasting feature in precise predicting difficulty, introducing more challenges to the distribution network operator (DSO).\par

With the increasing penetration of behind-the-meter distributed energy resources (DER), including rooftop solar photovoltaic (PV) systems, home battery storage, electric vehicles and flexible loads, the way of electricity generation and consumption is changing. While DER can reduce energy expenditure for the owners, they can also create network problems, including reverse power flows and overvoltages. In countries with a high rooftop solar PV penetration, low-voltage distribution networks are increasingly often at the verge of operating limits. Yet, distribution network operators (DSO) are ill-equipped to deal with the challenge. The issue is further exacerbated by the regulatory separation between network service provision and electricity retail and the partial smart meter rollout. In Australia, for example, only retailers have access to real-time smart meter data. They can sell the data to network companies, but only for the previous day. On top of that, the smart meter rollout is only complete in Victoria, while in other states, most customers still use accumulation meters. All these issues make active network management difficult. 

For the lack of better solutions, DSOs use blanket rooftop solar export limits; that is, they limit solar exports conservatively to ensure that network limits are not violated. However, this can lead to suboptimal outcomes. Against this backdrop, the authors in \cite{Petrou2020operating, petrou2021ensuring, Blackhall2020} proposed to use dynamic operating limits (also called operating envelopes) to limit PV exports. However, they assume full real-time observability of end-user demand and solar generation, which is unrealistic for the reasons explained above. To address the above issues, we proposed a chance-constrained AC optimal power flow (CC AC OPF) to compute operating envelopes under uncertainty \cite{Yu_PSCC_2022}. However, we assumed that the probability distributions of uncertain electricity demand and solar generation is known and given without going into details of how they can be obtained in practice. We address the gap in this paper by proposing a conditional generative adversarial network (CGAN)-based model to forecast household solar generation and electricity demand.

The remainder of the paper is described as follows: Section~$\text{\uppercase\expandafter{\romannumeral2}}$ reviews the literature of demand and solar generation forecasting. Section~$\text{\uppercase\expandafter{\romannumeral3}}$ formulates the CC AC OPF problem for the computation of dynamic operating limits. The forecasting methodology using the proposed conditional GAN is presented in Section~$\text{\uppercase\expandafter{\romannumeral4}}$. Section~$\text{\uppercase\expandafter{\romannumeral5}}$ presents simulation results, and Section~$\text{\uppercase\expandafter{\romannumeral6}}$ concludes the paper.

\section{Literature Review}
Forecasting information is the basis for the planning, management, and energy trading in electricity networks. Both the electricity demand and solar generation exhibit seasonal behavior. The prediction results at a certain time slot are related to not only the historical data at the previous time slot but also at the same time slot of the previous day or on the days in previous weeks \cite{hippert2001neural}. Moreover, short-term forecasting (seconds to minutes, hours, or days ahead) is challenged by some unexpected issues like accidental human behaviors or weather-related conditions \cite{ chaturvedi2016solar}. The stepping variations of load or energy output will weaken the operation economics and threaten system stability. Hence, it is essential for network managers to use and integrate the load and solar generation forecasts into power system optimal power flow problems. Various forecasting approaches and research improve the techniques to obtain the point forecast information. The approaches of which usually contain input classification, modal regression formulation, and predictions output\cite{negnevitsky2009overview}. A Long- and Short Time-series network (LSTNet) embedded with Convolution Neural Network (CNN) and the Recurrent Neural Network (RNN) is proposed in \cite{lai2018modeling}, which performs well on the points forecasts of both the load and energy generation. The LSTNet model can either discover local dependency patterns between multi-dimensional inputs or capture complex relevance in the long period. A spatiotemporal graph dictionary learning (STGDL) model is used in \cite{khodayar2020spatiotemporal} to generate the behind-the-meter forecasts of load and PV power. However, the deterministic forecasts are no longer sufficient in daily operations due to the variability and uncertainty of the energy generation, which increases the importance of the probabilistic forecasts. 

The probabilistic forecasts are usually indicated in the form of intervals, quantiles, or density functions. The forecasting interval is valuable for network operators to correct the prediction errors or inconsistencies in the day-ahead schedule \cite{hong2016probabilistic }. The conditional forecast residuals are defined in \cite{wang2018conditional} to integrate both point forecasts and quantile regression in the probabilistic forecast output. A constrained quantile regression averaging (CQRA) method is studied in \cite{wang2018combining} to find the optimal weights for the individual forecasting model to improve the performance of the combined models. Authors in \cite{feng2019reinforced} propose a reinforcement learning-based framework to choose the probabilistic forecasting model which has the highest performance dynamically. Probabilistic forecasts take the probability distributions to account for the variation information. Sometimes, in the process of solving stochastic optimization models, it is necessary to generate random scenarios with the given probability distribution of uncertainties. Then the GAN model is introduced in the probabilistic forecasts. Authors in \cite{jiang2018scenario} use the GAN-based model to generate long-term stochastic scenarios for renewable energy. To accurately capture the short-term variations of loads and energy generation, the conditional GAN (CGAN) model was formulated in \cite{chen2018unsupervised}, \cite{wang2020modeling} to output the probabilistic forecasts. \par

Inspired by the work above, we establish a CGAN-based framework to produce the probabilistic load and solar generation forecasts across different time periods. Then, the forecasting residuals are applied into the CC AC OPF model to generate the operating envelops. 

\begin{figure*}[!ht]
  \centering
  \includegraphics[width=6in]{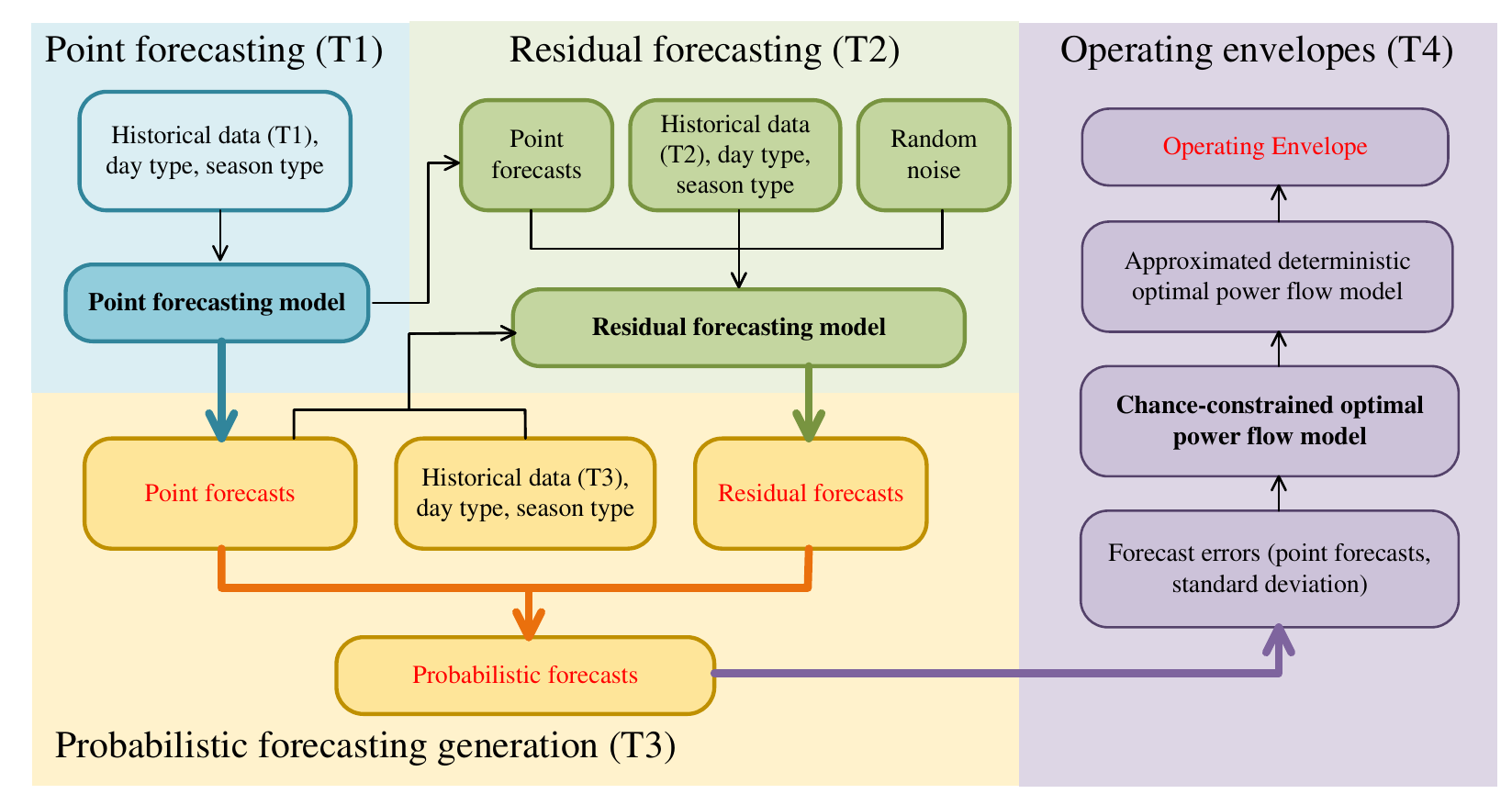} \\
  \caption{The proposed framework for the computation of operating envelopes under uncertainty. Blocks T1-T3 produce probabilistic forecasts of electricity demand and solar generation used in the chance-constrained optimal power flow in block T4. }         
  \label{fig:framework}
\end{figure*}

\section{Operating Envelopes under Uncertainty}
\subsection{Framework Overview}
Fig.~\ref{fig:framework} provides an overview of the computation of operating envelopes under uncertainty across three time horizons. A point forecasting model is trained first using a historical dataset (T1) and related factors such as day and season type to capture features relevant for the long-term forecast. Next, the historical dataset (T2) is used for testing and to compute forecasting residuals. After that, the historical dataset (T2) and related factors (day type, season type), point forecasts, forecasting residuals and random noise are fed into the conditional forecasting residual model. Thirdly, a new historical dataset (T3) is applied to the trained model in the last two steps. The point forecasts and point forecasting residuals constitute the probabilistic forecasts we need. Thus, the probabilistic forecasts (mean and standard deviation) are used to input the uncertain variables in the CC AC OPF model. Finally, we approximate the CC AC OPF model as a deterministic AC OPF model and produce the operating envelopes for individual prosumers.  

\subsection{Chance-constrained AC OPF Formulation}
Since household load and solar generation are difficult to forecast accurately, random forecast errors can adversely affect the computation of operating envelopes. To this end, we proposed a CC AC OPF model in our previous work \cite{Yu_PSCC_2022} to compute nodal exporting limitations under uncertainty. In our approach, the DSO computes nodal export limits using day-ahead probabilistic forecasts of solar generation and customer load and submits them to the customers as operating envelopes that serve as constraints in their optimisation. 

We consider a radial distribution network with a set of nodes $i \in \mathcal{N}$ and a set of lines $(i,j) \in \mathcal{L}$. For the sake of notational simplicity, we assume that one prosumer is connected to each node. The computation of operating envelopes is formulated as a chance-constrained OPF aiming to maximise network hosting capacity. We model the stochastic household demand $\tilde{p}^{\mathrm{d}}_{i}$ and solar generation $\tilde{p}^{\mathrm{pv}}_{i}$ as the sum of the point forecast part $\widehat{p}_{i}^{\mathrm{d},\mu}$ and $\widehat{p}_{i}^{\mathrm{pv},\mu}$ and forecasting residuals part $\varepsilon^{\mathrm{d}}_{i}$ and $\varepsilon^{\mathrm{pv}}_{i}$, respectively. The methodology to obtain the points forecasts and forecasting residuals is described in Section $\text{\uppercase\expandafter{\romannumeral4}}$. The chance constraints are introduced to guarantee that network constraints will not be violated when the inputs are subject to uncertainty. To ensure fair distribution of solar PV curtailment, we maximise the minimum export limit across all prosumers in a time slot $t \in \mathcal{T}$\footnote{The max-min formulation can be reformulated by introducing one auxiliary variable and a set of linear constraints for each $t \in \mathcal{T}$.}. 
We use a second-order cone (SOC) branch-flow OPF model, which results in the following probabilistic MISOCP OPF problem:
\begin{subequations} \label{Eq:DOPF_Obj}
	\begin{align}
		& \underset{x^{\text{p}} \in \mathcal{X}^{\text{p}}, \: x^{\text{n}} \in \mathcal{X}^{\text{n}}}{\text{max}} \sum_{t \in \mathcal{T}} \sum_{i \in \mathcal{N}} \underset{i} {\text{min}} \ p^{\mathrm{exp}}_{i,t}, \label{Eq:DSO_OPF_obj} \\	
		&\text{s.t.} \notag \\
		& \tilde{p}^{\text{pv}}_{i,t} \!\! - \!\! \tilde{p}^{\mathrm{d}}_{i,t} \!\! - \!\! \tilde{p}^{\mathrm{b}}_{i,t} \!\! = \!\! \sum_{j k \in \mathcal{L}} \!\! \left(\tilde{p}_{j k, t} \! + \! \tilde{\ell}_{j k, t} r_{j k}\right) \! - \!\!\sum_{i j \in \mathcal{L}} \!\! \tilde{p}_{i j, t}, \  i \! \in \mathcal{N}, \label{Eq:DOPF_Pbalance}\\
		& \tilde{p}^{\text{pv}}_{i,t} - \tilde{p}^{\mathrm{d}}_{i,t} - \tilde{p}^{\mathrm{b}}_{i,t} \le \overline{p}^{\mathrm{exp}}_{i,t}, \  i \in \mathcal{N}, \\
		& \tilde{q}^{\mathrm{pv}}_{i,t} \! - \tilde{q}^{\mathrm{d}}_{i,t} \! = \! \sum_{j k \in \mathcal{L}} \! \left(\tilde{q}_{j k, t} \!+ \! \tilde{\ell}_{j k, t} x_{j k} \! \right) - \! \sum_{i j \in \mathcal{L}} \! \tilde{q}_{i j, t}, \  i \in \! \mathcal{N},  \label{Eq:DOPF_Qbalance} \\
		& p^{\mathrm{exp}}_{i,t} \le p^{\mathrm{exp}}, \  i \in \mathcal{N}, \label{Eq:DOPF_Pexmax}\\
		& \tilde{v}_{i,t} \! - \! \tilde{v}_{j,t} \! = \! 2\left(r_{i j} \tilde{p}_{i j,t} \! + \! x_{i j} \tilde{q}_{i j,t}\right) \! - \! \left|z_{i j}\right|^{2} \! {\tilde{\ell}}_{i j,t}, \:  (i,j) \in \mathcal{L},  \label{Eq:DOPF_Vbalance} \\
		& \left\|\begin{array}{c}
			2 \tilde{p}_{ij,t} \\
			2 \tilde{q}_{ij,t} \\
			{\tilde{\ell}}_{ij,t}-\tilde{v}{i,t}
		\end{array}\right\|_{2} \leq {\tilde{\ell}}_{ij,t}+\tilde{v}_{i,t}, \  (i,j) \in \mathcal{L}, \label{Eq:DOPF_relax} \\
		& \tilde{p}_{ij,t}^{2}+\tilde{q}_{ij,t}^{2} \leq  \overline{s}_{ij}^{2}, \  (i,j) \in \mathcal{L}, \label{Eq:DOPF_S} \\
		&  0 \le \tilde{p}^{\mathrm{pv}}_{t} \leq \overline{p}^{\mathrm{pv}}_{t}, \label{Eq:DOPF_PV_P} \\
	    &  \underline{p}^{\mathrm{b}} \leq  b_{t}\tilde{p}^{\mathrm{b},+}_{t}+(1-b_{t})\tilde{p}^{\mathrm{b},-}_{t} \leq \overline{p}^{\mathrm{b}}, \label{Eq:DOPF_Pb_limit}\\
		&  {\tilde{e}^{\mathrm{b}}_{t}-\tilde{e}^{\mathrm{b}}_{t-1}} = \left( \eta b_{t}\tilde{p}^{\mathrm{b},+}_{t}- \frac{(1-b_{t})}{\eta}\tilde{p}^{\mathrm{b},-}_{t}\right)\Delta{t},  \label{Eq:DOPF_Pb_soc} \\
		&  \underline{e}^{\mathrm{b}} \leq \tilde{e}^{\mathrm{b}}_{t} \leq \overline{e}^{\mathrm{b}}, \label{Eq:DOPF_soc_limit} \\
		&  b_{t} \in \left\lbrace   0,1 \right\rbrace, \label{Eq:DOPF_binary} \\
		&\mathbb{P}\left(\widetilde{v}_{i} \leq \overline{v}_{i} \right) \geq 1-\xi^{\text{v}}, \label{Eq:CC_Umax}\\
        &\mathbb{P}\left(\widetilde{v}_{i} \geq \underline{v}_{i} \right) \geq 1-\xi^{\text{v}}, \label{Eq:CC_Umin}\\
        &\mathbb{P}\left(\widetilde{p}_{ij}^{2}+\widetilde{q}_{ij}^{2} \leq \overline{s}_{ij}^{2}\right) \geq 1-\xi^{\ell}, \label{Eq:CC_S}
\\ 
        &\mathbb{P}\left(\widetilde{p}_{i,t}^{\mathrm{b}} \leq \overline{p}^{\mathrm{b}}\right) \geq 1-\xi^{\text{p}}, \label{Eq:CC_Pbmax}\\
        &\mathbb{P}\left(\widetilde{p}_{i,t}^{\mathrm{b}} \geq \underline{p}^{\mathrm{b}}\right) \geq 1-\xi^{\text{p}}. \label{Eq:CC_Pbmin}
	\end{align}
\end{subequations}
where $\mathcal{X}^{\text{n}} = \left\lbrace p^{\mathrm{exp}}_{i,t}, \tilde{v}_{i,t}, \tilde{u}_{i,t}, \tilde{\ell}_{i,t}, \tilde{p}_{ij,t}, \tilde{q}_{ij,t} \right\rbrace $ for each $t \in \mathcal{T}$, $(i,j) \in \mathcal{L}$ and $i \in \mathcal{N}$ is the vector of network decision variables; $\mathcal{X}^{\text{p}} = \left\lbrace \tilde{p}^{\mathrm{pv}}_{i}, \tilde{p}^{\mathrm{b}}_{i,t}, \tilde{e}^{\mathrm{b}}_{t}, b_{t} \right\rbrace $ for each $t \in \mathcal{T}$, $(i,j) \in \mathcal{L}$ and $i \in \mathcal{N}$ for each $t \in \mathcal{T}$ is the vector of prosumer decision variables; $\tilde{\bullet}$ represents the stochastic variables.
The constraints include active and reactive power balance for each node \eqref{Eq:DOPF_Pbalance}-\eqref{Eq:DOPF_Qbalance}; maximum export limit \eqref{Eq:DOPF_Pexmax}; voltage drop \eqref{Eq:DOPF_Vbalance}, where $r_{i j}$ and $x_{i j}$ are resistance and reactance of the line, respectively, and $z_{i j}=r_{i j}+jx_{i j}$ is the line impedance; \eqref{Eq:DOPF_relax} is the SOC relaxation of the line power flow; \eqref{Eq:DOPF_S} is the line power flow limit; \eqref{Eq:DOPF_PV_P} - \eqref{Eq:DOPF_binary} are prosumers solar PV curtailment and battery constraints, respectively.
For reactive powers $\tilde{q}^{\mathrm{pv}}$ and $\tilde{q}^{\mathrm{d}}$ we assume a constant power factor $\text{cos}\varphi = 0.9$. \eqref{Eq:CC_Umax}-\eqref{Eq:CC_Pbmin} are stochastic constraints ensuring the probability of a stochastic variable $\widetilde{x}$ remaining within limits is $1-\xi^{\text{x}}$, where  \eqref{Eq:CC_Umax}-\eqref{Eq:CC_Umin} are the voltage magnitude limits, \eqref{Eq:CC_S} is the apparent power flow limit, and \eqref{Eq:CC_Pbmax}-\eqref{Eq:CC_Pbmin} ensures battery storage power limits. The CC AC OPF model is then reformulated into the deterministic AC OPF model referred to \cite{lubin2019chance}, which can be solved
efficiently.

\section{Demand and Generation Forecasting}
Solar generation and load forecasting are affected by various factors such as radiation, cloud coverage, temperature, user behavior, etc., and their interaction is intricate. No forecasting approach can capture all the influencing factors and generate absolute perfect results. Hence, the forecasting residuals are introduced to model the characteristics not captured in point forecasting. The point forecasts $\widehat{y_{t}}=f\left(\mathbf{X}_{t}\right)$ and forecasting residuals $\varepsilon_{t}$ are respectively considered as deterministic and uncertain part to form the probabilistic forecasts $y_{t}$. The input factor $\mathbf{X}_{t}$ contains the historical data, day type and season type. The forecasting residuals denotes the errors  between real and predicted values. In this paper, we use the long short-term memory network (LSTN) model to train and test for the point forecasts, and a CGAN model for forecasting the residuals. 

The whole forecasting process is conducted along three time horizons, with each using a different dataset. In the first step, the LSTN $f\left(\cdot\right)$ is trained using dataset T1. In the second step, we test the model $f\left(\mathbf{X}_{T_1}\right)$ with dataset T2 and calculate the forecast residuals $\varepsilon_{T_2}=y_{T_2}-f\left(\mathbf{X}_{T_2}\right)$. The CGAN model $g\left(\mathbf{C}_{t},z\right)$ is also trained, where $\mathbf{C}_{t}$ denotes condition factors (e.g. forecasts $y_{T_2}$, historical data T2, and day type) and $z$ denotes the random noise. The generator of the CGAN model can sample different residuals depending on the random noise $z$. In the third step, the point forecasting model $f\left(\mathbf{X}_{T_3}\right)$ and the residual forecasting model $g\left(\mathbf{C}_{T_3},z\right)$ use dataset T3 to generate scenario forecasts. The performance of the uncertainty and variation of the forecasts are usually evaluated the mean square error (MSE) or the mean absolute error (MAE). In this paper, the forecasts are evaluated in terms of the continuous ranked probability (CRPS) and the pinball loss (PL), which are more appropriate in our setting, as explained in Section~IV-B.

\begin{figure}[]
  \centering
  \includegraphics[width=3.5in]{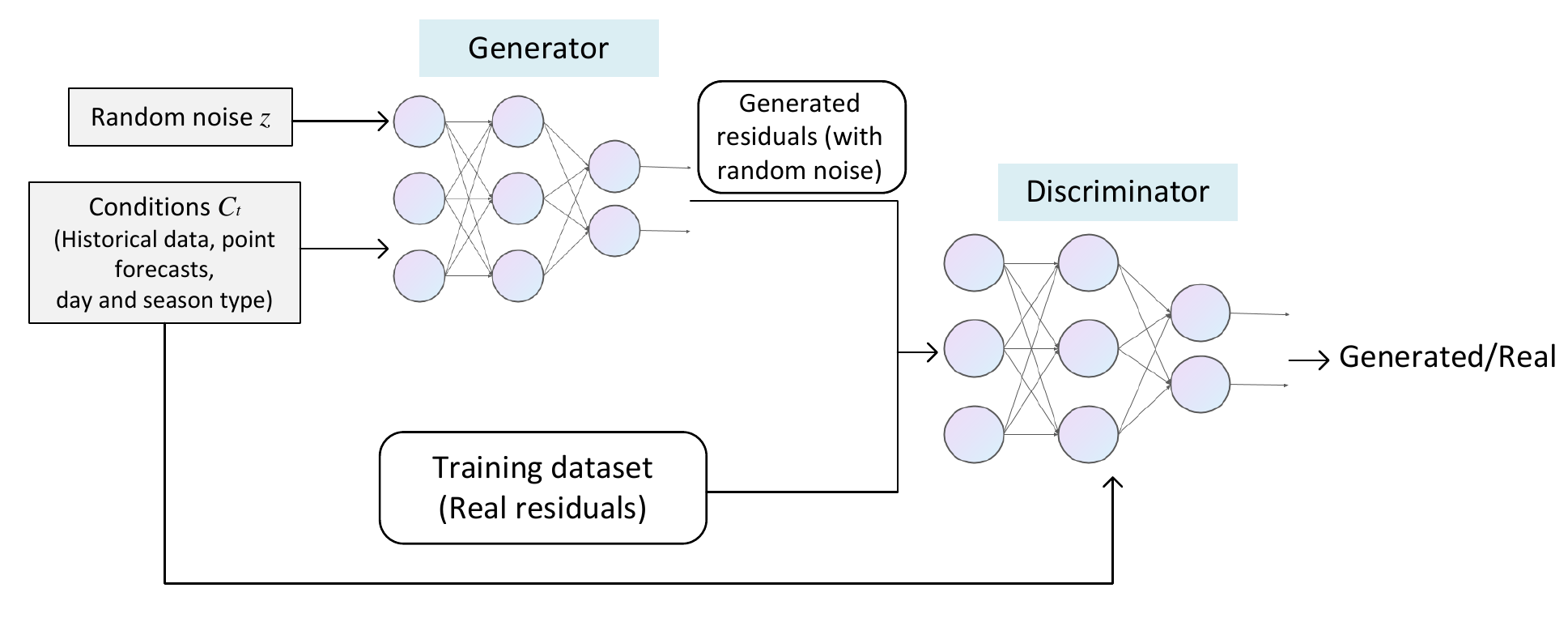} \\
  \caption{Structure of the conditional generative adversarial network (CGAN).}         
  \label{fig:CGAN}
\end{figure}

\subsection{Conditional Generative Adversarial Network (CGAN)}
The general task of the Generative Adversarial Network (GAN) model is to train two neural networks, called generator $G$ and discriminator $D$ in an adversarial learning mechanism. The CGAN model is the improvement of GAN model to capture variations comprehensively, where the condition $C$ is added as the inputs. The structure of the CGAN is shown in Fig.~\ref{fig:CGAN}, where the generator $g\left(\mathbf{C}_{T_3},z\right)$ generates a large number of forecasting residual scenarios $S_g$ with different random noise $z$ until they are infinitely close to the real dataset $S_r$ (the real forecasting residuals). The noise $z$ usually follows a pre-set distribution $z\sim\mathbb{P}_z$, such as Gaussian or uniform distribution. The discriminator compares the generated sample $S_g$ to the real ones to determine whether it should be accepted or rejected. The required parameters for generator and discriminators are $\theta_G$ and $\theta_D$, respectively. The sample vectors input to discriminator are represented as $S_g=G(z,\theta_{G})$. The discriminator $D(\cdot)$ distinguish whether samples are generated or real by estimating the values of the probability $p_{i}=D\left(G\left(z ; \theta_{G}\right) \mid \mathbf{C}, \theta_{D}\right)$. \par

The cross-entropy-based losses of generators and discriminators in the form of expectation for sample vector $S_g$ is defined as:
%\begin{small}
\begin{subequations} \label{Forecast_GD_loss}
\begin{align}
&L_{G}=\mathbb{E}_{Z}\left[\log \left(1-D\left(G\left(z ; \theta_{G}\right) \mid \mathbf{C}, \theta_{D}\right)\right)\right], \label{Forecast_G_loss}\\
\begin{split}
	L_{D}=-\mathbb{E}_{S_{r}}\left[\log \left(D\left(s_{r} \mid \mathbf{C} ; \theta_{D}\right)\right)\right]\\
	-\mathbb{E}_{Z}\left[\log \left(1-D\left(s_{g} \mid \mathbf{C} ; \theta_{D}\right)\right)\right] \label{Forecast_D_loss},
\end{split}
\end{align}
\end{subequations}
%\end{small}
where $\mathbb{E}_{Z}$ and $\mathbb{E}_{S_r}$ represent generated sample inputs and the expected values among real sample inputs, receptively. During training, the generator and discriminator are optimized simultaneously. The generated samples get more  similar to the real ones with the iteration number increasing.  
\par
The adversarial mechanism between two neural networks $G$ and $D$ is formulated as a min-max optimization problem with the following objective:
\begin{equation} \label{Forecast_GD_obj}
\begin{split}
\min _{\theta_{G}} \max _{\theta_{D}} &\mathbb{E}_{S_{r}}\left[\log \left(D\left(s_{r} \mid \mathbf{C} ; \theta_{D}\right)\right)\right] +\\
&\mathbb{E}_{Z}\left[\log \left(1-D\left(G\left(z ; \theta_{G}\right) \mid \mathbf{C}, \theta_{D}\right)\right)\right].
\end{split}
\end{equation} 
The authors in \cite{wang2020modeling} improved the GAN model to the Wasserstein GAN (WGAN) by adding a gradient penalty to accelerate the computation while stabilizing the training process and ensuring that the convergence rates of the generator and discriminator are synchronized.
Compared to GAN, the forecasting model in WGAN uses a linearized activation function in the last layer of the neutral network $D$, so that the weights of network $D$ are limited into a fixed range after each training. Finally, the logarithmic functions in the generator and discriminator loss functions are replaced by linear ones since gradients of logarithmic functions are negligible. Hence, the simplified loss functions can be rewritten as:
\begin{subequations} \label{Forecast_GD_loss_new}
\begin{align}
&L_{G}=-\mathbb{E}_{Z}\left[D\left(G\left(z ; \theta_{G}\right) \mid \mathbf{C}, \theta_{D}\right)\right], \\
&L_{D}=-\mathbb{E}_{S_{r}}\left[\left(D\left(s_{r} \mid \mathbf{C}, \theta_{D}\right)\right)\right]+\mathbb{E}_{Z}\left[D\left(s_{g} \mid \mathbf{C}, \theta_{D}\right)\right]
\end{align}
\end{subequations}
The gradient penalty ($GP$) is added to the discriminator loss function to make sure the distinguishing ability of discriminator under simplification, which is defined as:
\begin{equation}
G P=\lambda \mathbb{E}_{S}\left[\|\nabla D(\hat{s})\|_{2}-1\right]^{2},
\end{equation}
where $\lambda$ represents the hyper-parameter that determines the weight of $GP$; $\hat{s}=\rho S_g +\left(1-\rho\right) S_r$ represents the average of the generated samples $S_g$ and real data samples $S_r$, where $\rho$ is a random variable following uniform distribution $U(0,1)$; $\mathbb{E}_{S}$ represents the expected value among all $\hat{s}$. With the gradient penalty, the discriminator loss function is reformulated as:
\begin{equation}
L_{D}=-\mathbb{E}_{S_{r}}\left[\left(D\left(s_{r} \mid \mathbf{C}, \theta_{D}\right)\right)\right]+\mathbb{E}_{Z}\left[D\left(s_{g} \mid \mathbf{C}, \theta_{D}\right)\right]+G P.
\end{equation}
\subsection{Evaluation Criteria}
To evaluate stochastic forecasts, the continuous ranked probability (CRPS) and pinball loss (PL) are introduced in the probabilistic domain. The calculation of CRPS is formulated as:
\begin{equation}
\operatorname{CRPS}\left(F, y_{t}\right)=\int_{-\infty}^{\infty}\left(F\left(y_{t}^{\prime}\right)-1\left(y_{t}^{\prime}-y_{t}\right)\right)^{2} d y_{t}^{\prime},
\end{equation}
where $F\left(y_{t}^{\prime}\right)$ denotes the cumulative distribution function of the predicted variable at time $t$; $1\left(y_{t}^{\prime}-y_{t}\right)$ is the term to compare the values between $y_{t}^{\prime}$ and $y_t$, and the value of it equals to 1 if $y_{t}^{\prime} \textgreater y_t$ and equals to 0 otherwise. The forecasts are better with smaller CRPS. \par
The PL is capable of calculating both positive and negative errors for different quantile compared to MAE, which is defined as:
\begin{equation}
\operatorname{PL}\left(\hat{y}_{t, q}, y_{t}, q\right)= \begin{cases}(1-q)\left(\hat{y}_{t, q}-y_{t}\right) & y_{t}<\hat{y}_{t, q} \\ q\left(y_{t}-\hat{y}_{t, q}\right) & y_{t} \geq \hat{y}_{t, q}\end{cases},
\end{equation}
where $q$ denotes the quantile; $\hat{y}_{t, q}$ denotes the predicted quantile at time $t$.  

\section{Case study}
\subsection{Data Implementation}
The condition inputs $\mathbf{C}$ include historical data of six days before the forecasting day (1, 2, 3, 5, 7, 14 and 21 days before), the day type (day of the week $\textit{W}_d$) and season type (day of the season $\textit{S}_d$). Note the spring in the Southern Hemisphere begins in September. Hence, the condition input vector is
\begin{equation}
\mathbf{C}=\left[\mathbf{L}_{d-1}, \mathbf{L}_{d-2}, \mathbf{L}_{d-3}, \mathbf{L}_{d-7}, \mathbf{L}_{d-14}, \mathbf{L}_{d-21}, W_{d}, S_{d}\right],
\end{equation} 
where $\mathbf{L}_d-i$ represents the daily real data of $i$ days before the forecasting day $d$. We assume 30-min dispatch, resulting in the dimension of the condition input vector for one day to be 290 ($6 \times 2 \times 24 + 2 = 290$).

We use Ausgrid's Solar Home Electricity dataset\footnote{www.ausgrid.com.au} consisting of solar generation and household electricity demand from 2010-2012. Dataset T1, between 1 July 2010 to 30 June 2011, is used to train the LSTN model, while dataset T2, from 1 July 2011 to 30 June 2012 (except 7th, 14th and 28th of each month, 36 days in total) is used to test the point forecasts and to train the CGAN model. The testing dataset T3, used to  evaluate the performance of the forecasting framework, consists of the days that were excluded from the training dataset T2 (7th, 14th and 28th day of each month, 36 days in total).
 
\subsection{Parameter Setup}
We use a three-layer network for both generator and discriminator, containing an input layer, a hidden layer, and an output layer. The number of neurons in the input layer of generator is $802=290+512$. The hidden layer has 256 neurons while the output layer generate 48 scenarios. The dimension of random noise $z$ is 512, which is larger than the condition input $\mathbf{C}$. For the discriminator, there are 338 neurons in the input layer (48 for the number of samples, 290 for the condition input vector). The number of neurons in the hidden layer is 128 and the output layer has one neuron. 
The activation functions of the hidden and the output layer are ReLU and Sigmoid, respectively. The training strategy for the CGAN model is to train the generator once and the discriminator five times, which balances the coverage speed of two neurons network. The simulations were produced using the optimisation modelling toolbox YALMIP with CPLEX solver in MATLAB® 2020 on a 64-bit laptop with a 2.60 GHz CPU and 16GB RAM.

\subsection{Neural Network training}
\begin{figure}[]
  \centering
  \includegraphics[width=3.25in]{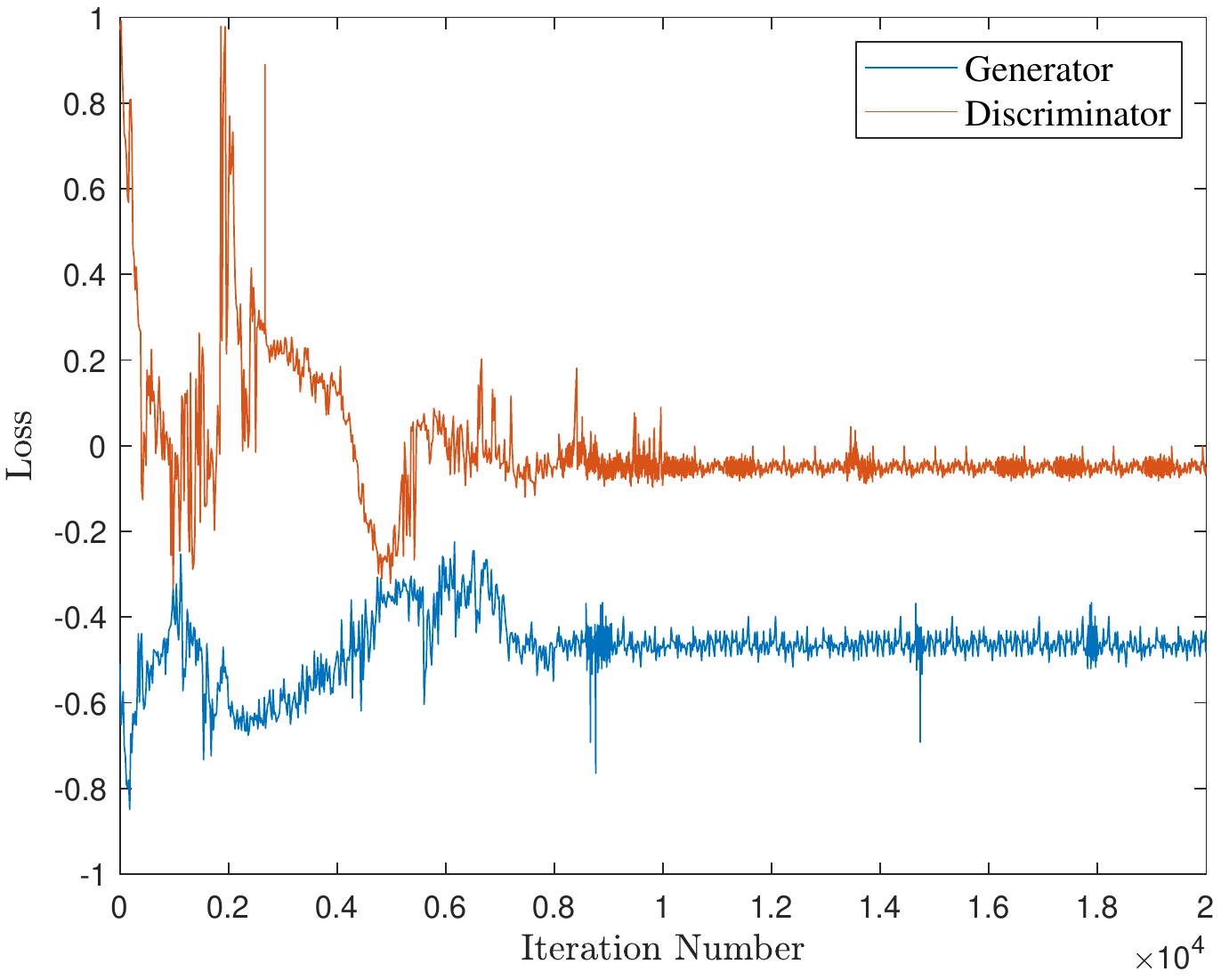} \\
  \caption{Losses of generator and discriminator}         
  \label{fig:loss}
\end{figure}

Fig.~\ref{fig:loss} illustrates the learning process showing the loss functions of the generator and discriminator over 20,000 iterations. Observe that the training for both converges after about 7,000 iterations. The convergence value of the discriminators is near 0, which means the sample scenarios generated by generators is well approached to the real residuals. The CRPS are 27.38 and 48.62 for load forecasts and solar generation forecasts, respectively, and the PL are 15.97 and 24.11 for load and solar generation forecasts, respectively, which shows that the generated scenarios capture the variation characteristics well for household load.    

\subsection{Forecasting Results}
\begin{figure}[]
  \centering
  \includegraphics[width=3.25in]{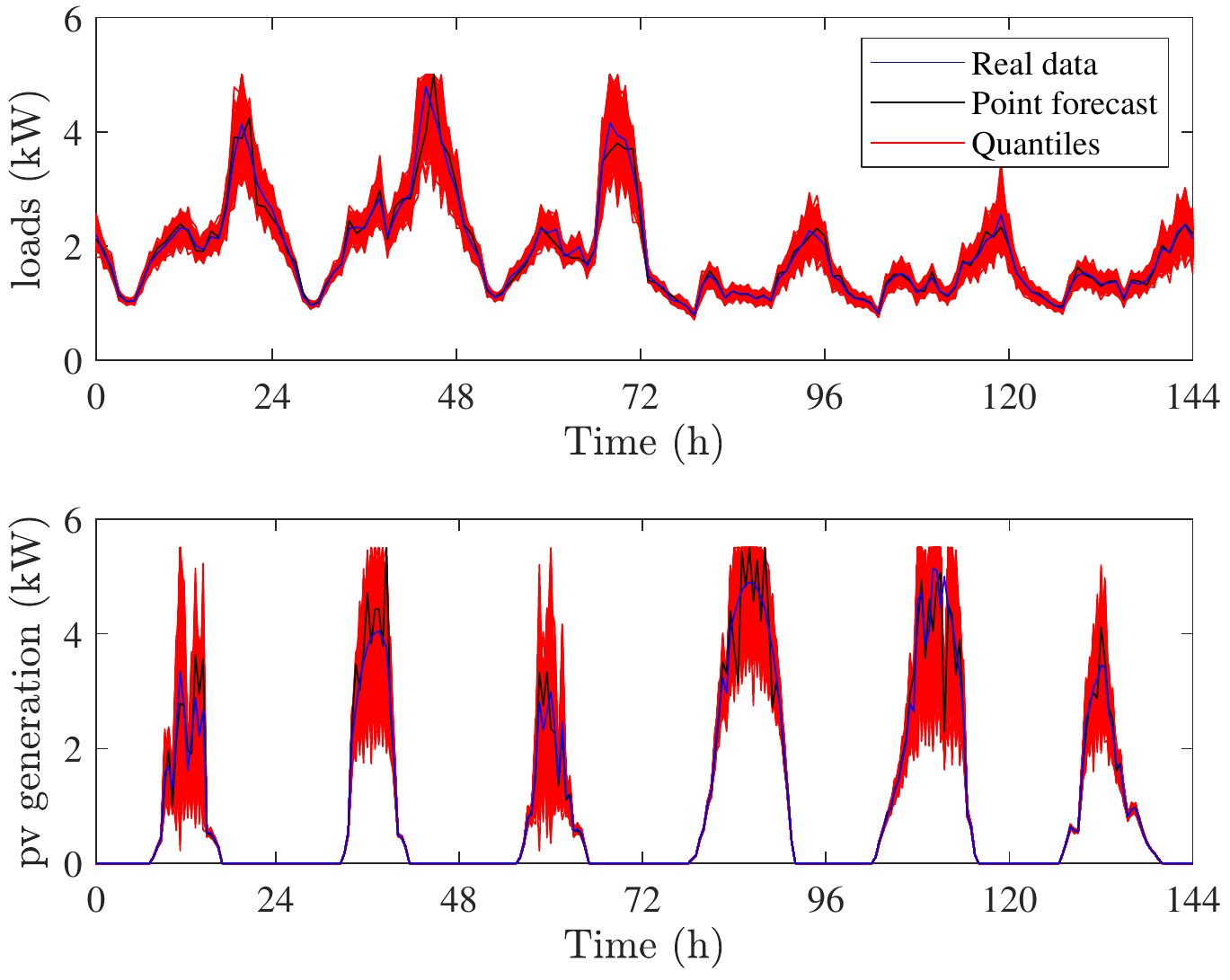} \\
  \caption{Scenarios generated with CGAN model.}         
  \label{fig:forecasts}
\end{figure}

Fig.~\ref{fig:forecasts} shows the real data, point forecasts and scenarios generated from generators of uncertain variables (loads and solar generation) at the day of 7th, 14th and 28th on July and December, respectively. 
We select the days of different to explore the regularity of predictions and forecasting residuals. 
Observe that the variation of solar generation forecasting residuals is larger than the one of the electricity demand.
This is because of the natural variability of solar insolation, which is affected by weather conditions, in particular the intermittent cloud coverage. In contrast, the impact of customer behavior on electricity demand forecast is less significant.
Observe also that the forecasting residuals are conditional on the time periods (e.g. the variations of electricity demand forecasting residuals is more significant at night while the solar generation forecasting residuals vary at daytime.)

\begin{figure}[]
  \centering
  \includegraphics[width=3.25in]{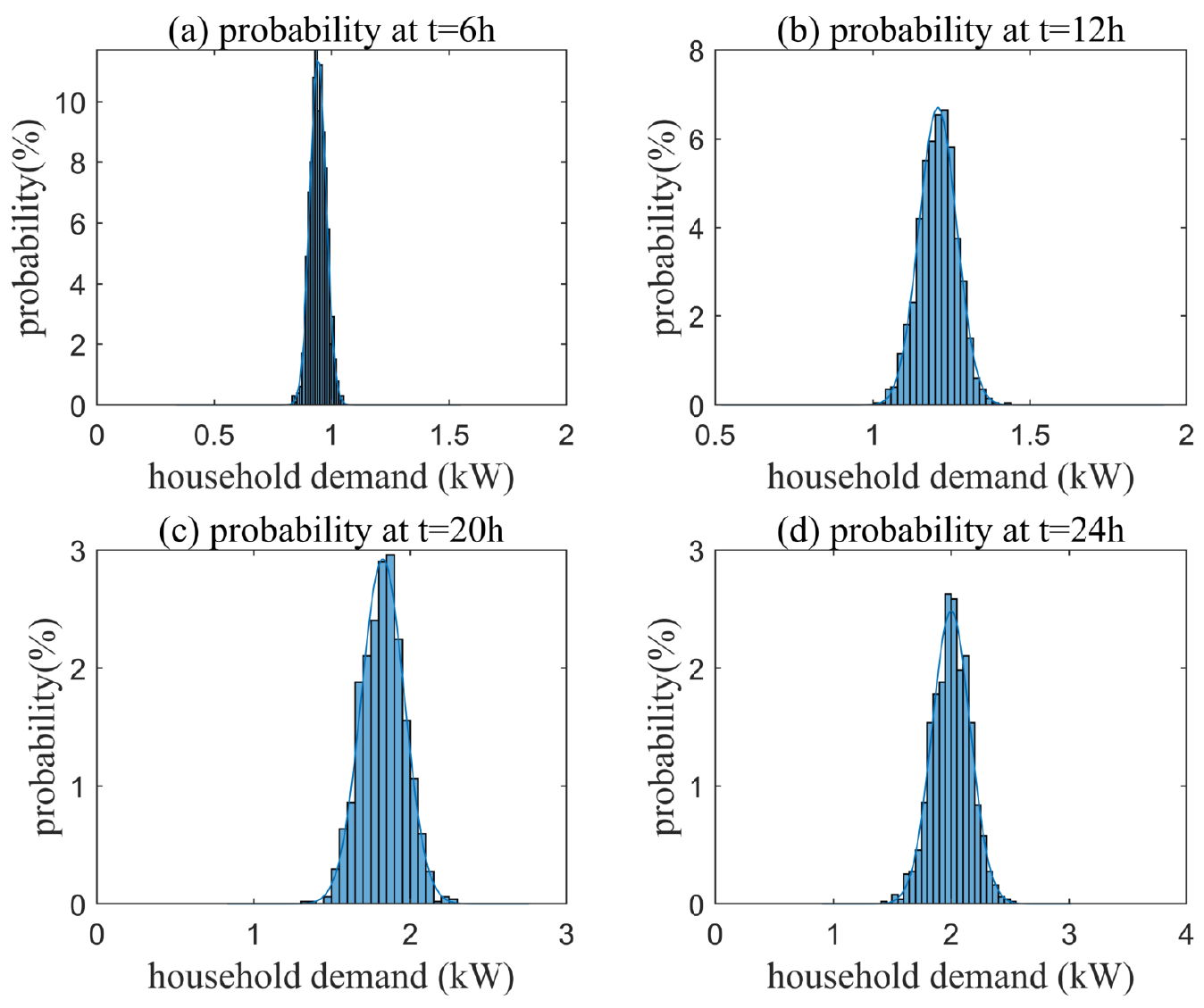} \\
  \caption{Probability distributions of electricity demand forecasts at different time periods.}         
  \label{fig:load_pro}
\end{figure}

\begin{figure}[]
  \centering
  \includegraphics[width=3.25in]{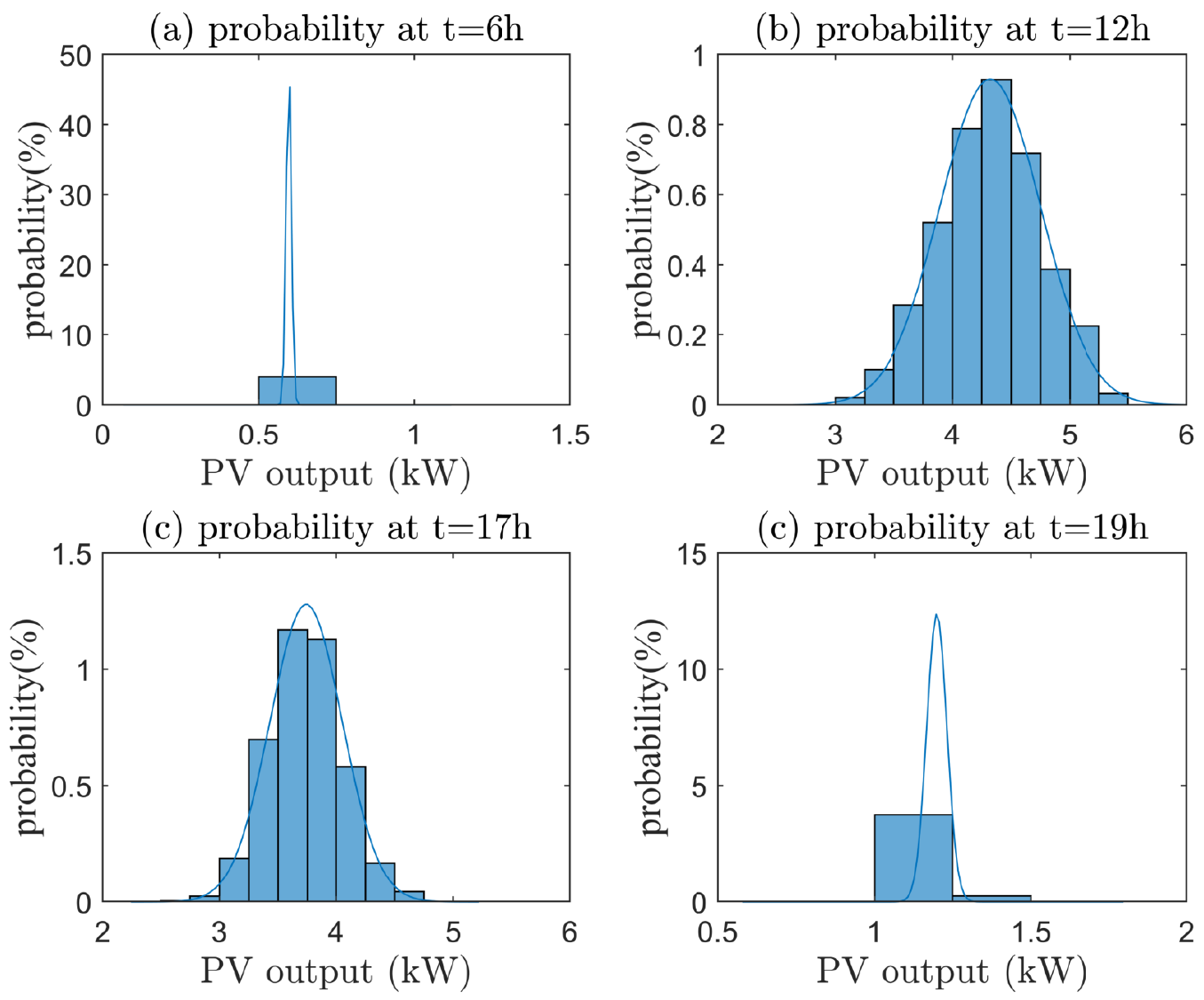} \\
  \caption{Probability distribution of solar generation forecasts at different time periods}         
  \label{fig:lpv_pro}
\end{figure}

The Gaussian probability distributions of electricity demand and solar generation forecasts at different time periods are shown in Figs.~\ref{fig:load_pro} and \ref{fig:lpv_pro}, respectively. Observe that the electricity demand forecast is more volatile in the afternoon and evening due to the increased customer activity when they return from work, and less volatile in the morning and in the middle of the day. By contrast, the solar generation forecast is more volatile during the day due to the weather conditions,  in particular the intermittent cloud coverage.

\subsection{Operating Envelopes under Probabilistic Forecasts}
\begin{figure}[]
  \centering
  \includegraphics[width=3.25in]{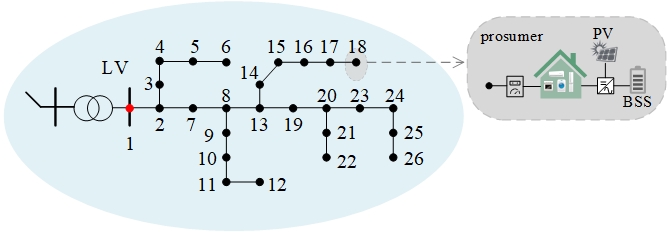} \\
  \caption{Distribution network topology.}         
  \label{fig:network}
\end{figure}

We use a distribution network test system with 25 prosumers shown in Fig.~\ref{fig:network} with HEMS, PV and battery storage systems (BSS) communicating with the DSO through a smart meter. The Time-of-use (ToU) electricity tariff and feed-in-tariff (FiT) are shown in Tab.~\ref{prices}. The voltage limitation is $\pm 6\%$ of the normalized values. Each prosumer is equipped with a \SI{6}{\kilo\watt}/\SI{14}{\kilo\watt\hour} PV-battery system. The battery charging/discharging limit is \SI{-3.5}{\kilo\watt}/\SI{3.5}{\kilo\watt}, and the energy capacity is limited between \SI{4}{\kilo\watt\hour} and \SI{10}{\kilo\watt\hour}. We fit a Gaussian distribution using the generated scenarios to obtain standard deviations of the electricity demand and solar generation in each time period. The parameters of the distribution are used to approximate the stochastic constraints in \eqref{Eq:DOPF_Obj}. 

\begin{figure}[!ht]
  \centering
  \includegraphics[width=3.25in]{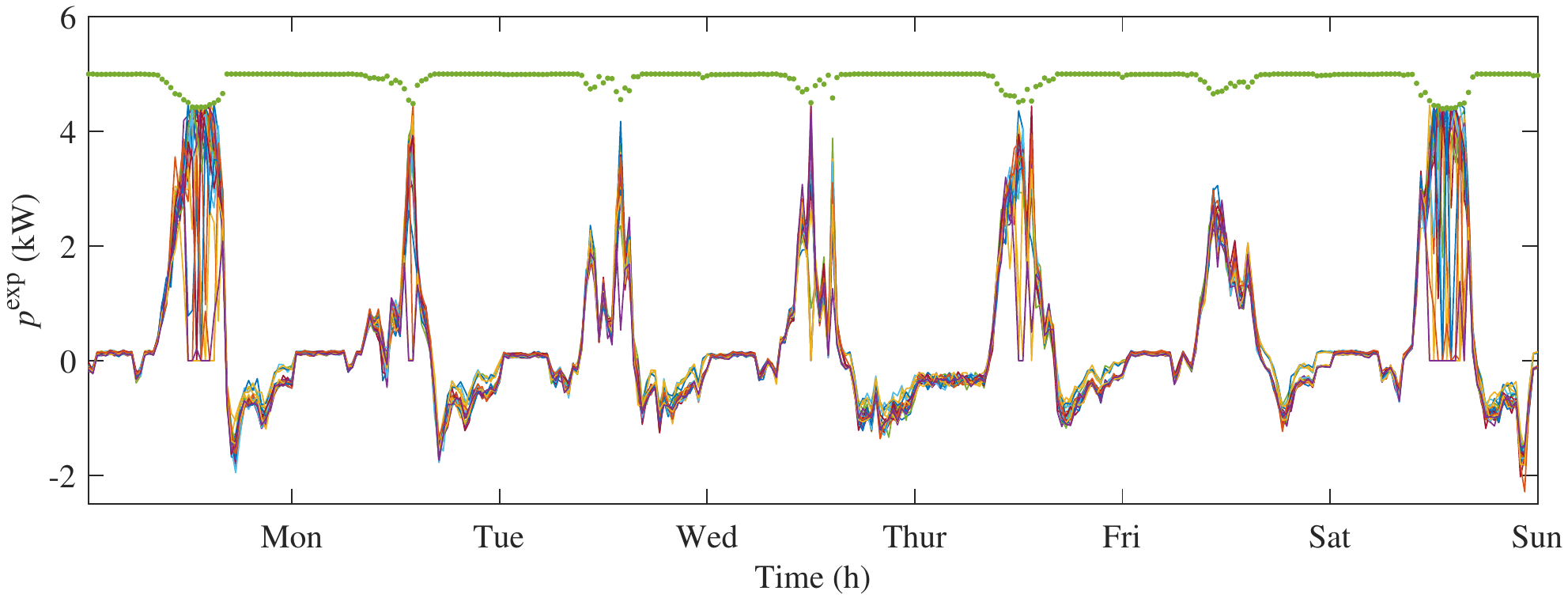} \\
  \caption{Weekly operating envelopes and average nodal power flows calculated by CC AF OPF model with probabilistic forecasts.}         
  \label{fig:OE_case_week}
\end{figure}

Fig.~\ref{fig:OE_case_week} shows the weekly operating envelopes and mean nodal power flows with different days of solar generation and demand forecasts. The nodal exported power is limited with regard to the forecasting residuals, ensuring the operations by DSO under uncertainties. The nodal exporting limitations are obvious on the days with high solar generation, but the exporting limitations relaxed on the days with low solar generation. The user-side solar PV curtailment increases at the middle of the day as the PV system provide more electricity. 
 
\begin{table}
\renewcommand{\arraystretch}{1.3}
\centering
\caption{Time-of-use (ToU) electricity tariff and feed-in-tariff (FiT).}
\label{prices}
\begin{tabular}{cccccc}
\hline
Time slot h) & 0-7 & 7-15 & 15-21 & 21-22 & 22-24 \\
\hline
ToU(c/kWh) & 15.96 & 25.96 & 57.76 & 25.96 & 15.96\\
\hline
FiT(c/kWh) & 9 & 9 & 9 & 9 & 9\\
\hline
\end{tabular}
\end{table}

\section{Conclusion}
This paper has proposed a novel methodology for computing operating envelopes under probabilistic electricity demand and solar generation forecasts. The probability distributions of electricity demand and solar generation is captured by a CGAN-based model, which is conditional on the historical dataset, point forecasts and time horizons. The parameters of the Gaussian distribution (mean and standard deviations) are calculated using the generated residual scenarios, which is applied in the CC AC OPF problem to limit the nodal operating envelopes. The effectiveness and performance of our proposed framework is evaluated by the case studies.

\bibliographystyle{IEEEtran}
\bibliography{yuyi_IREP}

\begin{thebibliography}{10}
\providecommand{\url}[1]{#1}
\csname url@samestyle\endcsname
\providecommand{\newblock}{\relax}
\providecommand{\bibinfo}[2]{#2}
\providecommand{\BIBentrySTDinterwordspacing}{\spaceskip=0pt\relax}
\providecommand{\BIBentryALTinterwordstretchfactor}{4}
\providecommand{\BIBentryALTinterwordspacing}{\spaceskip=\fontdimen2\font plus
\BIBentryALTinterwordstretchfactor\fontdimen3\font minus
  \fontdimen4\font\relax}
\providecommand{\BIBforeignlanguage}[2]{{%
\expandafter\ifx\csname l@#1\endcsname\relax
\typeout{** WARNING: IEEEtran.bst: No hyphenation pattern has been}%
\typeout{** loaded for the language `#1'. Using the pattern for}%
\typeout{** the default language instead.}%
\else
\language=\csname l@#1\endcsname
\fi
#2}}
\providecommand{\BIBdecl}{\relax}
\BIBdecl

\bibitem{Petrou2020operating}
K.~Petrou, M.~Z. Liu, A.~T. Procopiou, L.~F. Ochoa, J.~Theunissen, and
  J.~Harding, ``{Operating envelopes for prosumers in LV networks: A weighted
  proportional fairness approach},'' in \emph{IEEE PES Innovative Smart Grid
  Technologies Conference Europe}, 2020.

\bibitem{petrou2021ensuring}
K.~Petrou, A.~T. Procopiou, L.~Gutierrez-Lagos, M.~Z. Liu, L.~F. Ochoa,
  T.~Langstaff, and J.~Theunissen, ``{Ensuring distribution network integrity
  using dynamic operating limits for prosumers},'' \emph{IEEE Trans. Smart
  Grid.}, 2021.

\bibitem{Blackhall2020}
L.~Blackhall, ``{On the calculation and use of dynamic operating envelopes,
  Evolve Project M4 Knowledge Sharing Report},'' in \emph{Technology report},
  2020.

\bibitem{Yu_PSCC_2022}
Y.~Yi and G.~Verbi\v{c}, ``Fair operating envelopes under uncertainty using
  chance constrained optimal power flow,'' in \emph{2022 Power Systems
  Computation Conference (PSCC)}, 2022.

\bibitem{hippert2001neural}
H.~S. Hippert, C.~E. Pedreira, and R.~C. Souza, ``Neural networks for
  short-term load forecasting: A review and evaluation,'' \emph{IEEE
  Transactions on power systems}, vol.~16, no.~1, pp. 44--55, 2001.

\bibitem{chaturvedi2016solar}
D.~Chaturvedi and I.~Isha, ``Solar power forecasting: A review,''
  \emph{International Journal of Computer Applications}, vol. 145, no.~6, pp.
  28--50, 2016.

\bibitem{negnevitsky2009overview}
M.~Negnevitsky, P.~Mandal, and A.~K. Srivastava, ``An overview of forecasting
  problems and techniques in power systems,'' in \emph{2009 IEEE Power \&
  Energy Society General Meeting}.\hskip 1em plus 0.5em minus 0.4em\relax IEEE,
  2009, pp. 1--4.

\bibitem{lai2018modeling}
G.~Lai, W.-C. Chang, Y.~Yang, and H.~Liu, ``Modeling long-and short-term
  temporal patterns with deep neural networks,'' in \emph{The 41st
  International ACM SIGIR Conference on Research \& Development in Information
  Retrieval}, 2018, pp. 95--104.

\bibitem{khodayar2020spatiotemporal}
M.~Khodayar, G.~Liu, J.~Wang, O.~Kaynak, and M.~E. Khodayar, ``Spatiotemporal
  behind-the-meter load and pv power forecasting via deep graph dictionary
  learning,'' \emph{IEEE transactions on neural networks and learning systems},
  2020.

\bibitem{hong2016probabilistic}
T.~Hong and S.~Fan, ``Probabilistic electric load forecasting: A tutorial
  review,'' \emph{International Journal of Forecasting}, vol.~32, no.~3, pp.
  914--938, 2016.

\bibitem{wang2018conditional}
Y.~Wang, Q.~Chen, N.~Zhang, and Y.~Wang, ``Conditional residual modeling for
  probabilistic load forecasting,'' \emph{IEEE Transactions on Power Systems},
  vol.~33, no.~6, pp. 7327--7330, 2018.

\bibitem{wang2018combining}
Y.~Wang, N.~Zhang, Y.~Tan, T.~Hong, D.~S. Kirschen, and C.~Kang, ``Combining
  probabilistic load forecasts,'' \emph{IEEE Transactions on Smart Grid},
  vol.~10, no.~4, pp. 3664--3674, 2018.

\bibitem{feng2019reinforced}
C.~Feng, M.~Sun, and J.~Zhang, ``Reinforced deterministic and probabilistic
  load forecasting via $ q $-learning dynamic model selection,'' \emph{IEEE
  Transactions on Smart Grid}, vol.~11, no.~2, pp. 1377--1386, 2019.

\bibitem{jiang2018scenario}
C.~Jiang, Y.~Mao, Y.~Chai, M.~Yu, and S.~Tao, ``Scenario generation for wind
  power using improved generative adversarial networks,'' \emph{IEEE Access},
  vol.~6, pp. 62\,193--62\,203, 2018.

\bibitem{chen2018unsupervised}
Y.~Chen, X.~Wang, and B.~Zhang, ``An unsupervised deep learning approach for
  scenario forecasts,'' in \emph{2018 Power Systems Computation Conference
  (PSCC)}.\hskip 1em plus 0.5em minus 0.4em\relax IEEE, 2018, pp. 1--7.

\bibitem{wang2020modeling}
Y.~Wang, G.~Hug, Z.~Liu, and N.~Zhang, ``Modeling load forecast uncertainty
  using generative adversarial networks,'' \emph{Electric Power Systems
  Research}, vol. 189, p. 106732, 2020.

\bibitem{lubin2019chance}
M.~Lubin, Y.~Dvorkin, and L.~Roald, ``Chance constraints for improving the
  security of ac optimal power flow,'' \emph{IEEE Transactions on Power
  Systems}, vol.~34, no.~3, pp. 1908--1917, 2019.

\end{thebibliography}

\end{document}